# Kinetics and thermodynamics of carbon segregation and graphene growth on Ru(0001)

Kevin F. McCarty,[*,1] Peter J. Feibelman,[2] Elena Loginova,[1] and Norman C. Bartelt[1]

[1]Sandia National Laboratories
Livermore, CA 94550

[2]Sandia National Laboratories
Albuquerque, NM 87185

**Abstract**

We measure the concentration of carbon adatoms on the Ru(0001) surface that are in equilibrium with C atoms in the crystal's bulk by monitoring the electron reflectivity of the surface while imaging. During cooling from high temperature, C atoms segregate to the Ru surface, causing graphene islands to nucleate. Using low-energy electron microscopy (LEEM), we measure the growth rate of individual graphene islands and, simultaneously, the local concentration of C adatoms on the surface. We find that graphene growth is fed by the supersaturated, two-dimensional gas of C adatoms rather than by direct exchange between the bulk C and the graphene. At long times, the rate at which C diffuses from the bulk to the surface controls the graphene growth rate. The competition among C in three states – dissolved in Ru, as an adatom, and in graphene – is quantified and discussed. The adatom segregation enthalpy determined by applying the simple Langmuir-McLean model to the temperature-dependent equilibrium concentration seriously disagrees with the value calculated from first-principles. This discrepancy suggests that the assumption in the model of non-interacting C is not valid.

**1. Introduction**

Surface scientists have often focused attention on surface carbon, and on the phenomenon of surface segregation. Carbon can be an unwanted impurity segregated from the bulk of a material to its surfaces [1], a key element of industrially important catalytic chemistries [2], and a component for materials synthesis (e.g., synthetic diamond [3] and silicon carbide [4].) Most recently, the electronic properties of single sheets of graphitic carbon, graphene, have received considerable attention [5-8].

The C source in catalysis and material synthesis is typically a vapor-phase species, e.g., a hydrocarbon, but carbon from a material's bulk can also participate in surface reactions. This bulk C can originate from vapor-phase C sources. For example, C from vapor-phase sources that is transported through the bulk of substrates and catalysts is thought to be important to C fiber and nanotube synthesis [9, 10]. As temperature is reduced, the bulk solubility of C as an impurity in materials is reduced. Thus, there is a tendency of C to segregate to surfaces with cooling [11]. If the amount of bulk C is sufficiently high, a single layer of graphene can nucleate and grow from C segregating on metals during cooling [12]. At higher bulk concentrations, multilayers of C can precipitate [13-16]. A basic and unanswered question is whether this condensed C forms directly from C in the bulk (in an Eley-Rideal-like mechanism) or through an intermediate, adsorbed surface phase (a la Langmuir-Hinshelwood) [17].

To address this question, we study the kinetics and thermodynamics of C segregating from bulk Ru metal to its (0001) surface. C on this surface has been analyzed by theory [18, 19] and a variety of experimental techniques [14, 20-29]. We measure the coverage of C adatoms as a function of temperature by observing electron reflectivity [30] in a low-energy electron microscope (LEEM) [31]. Because our approach is imaging based, we can measure local adatom concentrations without being confounded by any C present as a condensed phase, i.e., graphene. At high temperature, before graphene nucleates, we can measure the adatom concentration in equilibrium with the bulk. The temperature dependence of this concentration yields a segregation enthalpy, which we compare with theory.

To determine the mechanism by which segregated C forms graphene, we quantify the competition between C adatoms in equilibrium with two C reservoirs -- C in the bulk of Ru and C in graphene [32]. We find that graphene nucleates and grows from C in the two-dimensional (2D) adatom gas rather than directly from C in the bulk. A supersaturation of adatoms fuels the initial graphene growth, and the growth rate slows as the supersaturation decreases. Then the graphene grows at the rate at which C diffuses through the Ru bulk to the surface, although growth still occurs through the intermediate surface 2D gas.

## 2. Experimental and computational details

Two Ru(0001) crystals were analyzed. The first was used to study C segregation, and contained enough bulk C [33] to cause measurable segregation when cooled to about 700–950°C. The second crystal was used to determine the concentration of C adatoms in equilibrium with graphene [28]. This crystal was thoroughly depleted of bulk C by repeatedly segregating the C into graphene islands (over hours to tens of hours) and then removing the C by exposing to oxygen at the segregation temperature. To measure the adatom/graphene equilibrium, C was deposited onto the Ru surface from a C rod heated by an electron beam, as previously described [28]. Temperatures were measured by thermocouples spot-welded to the side of the crystals.

All measurements were performed in an ultrahigh-vacuum-based LEEM. Auger spectroscopy confirmed that the segregating material was indeed C. Low-energy electron diffraction (LEED) confirmed the islands were graphene in an epitaxial relation to the Ru substrate (see Fig. 1, inset). The local electron reflectivity was measured from the intensity of LEEM images formed from the specularly reflected electron beam. When graphene islands were present, the reflectivity was measured from graphene-free regions, typically several microns in linear dimension (see Fig. 4).[1] The segregation experiments used a digitally compressed signal from an analog CCD camera to measure the reflectivity of 3.6-eV electrons, near the maximum decrease in reflectivity (see Fig. 1). We have previously established that the electron reflectivity decreases linearly with C adatom coverage [28]. To convert the reflectivity change to an absolute concentration of C adatoms, we use the relationship, adatom coverage = $0.223 \times [I_0 - I(t)]/I_0$ monolayers (ML), where $I_0$ is the image intensity from the local Ru region without C and $I(t)$ is the intensity at time $t$. Here, 1 ML equals the areal density of C in epitaxial graphene on Ru(0001). This calibration comes from our study of adatom/graphene equilibrium, which used an uncompressed digital CCD camera and 3.7-eV electrons [28]. Given the slightly different electron energy and different imaging system used for the segregation study, the concentrations of segregated C adatoms we report here are less accurate than those reported in our adatom/graphene equilibrium study [28].

The energy differences between C adatoms on the Ru(0001) surface and interior sites were determined through calculations similar to those of Ref. [28], using VASP [34, 35] in the PW91 generalized gradient approximation [36, 37] with electron–core interactions treated in the projector augmented wave approximation [38, 39]. Slabs of 7 Ru layers were simulated. Interstitials between layers 1 and 2 were calculated with the bottom 3 slab layers fixed in theoretical, bulk Ru relative positions. Interstitials between layers 3 and 4 were calculated with the bottom 2 layers fixed. The calculations were performed in 2√3 × 2√3 surface supercells, such that nearest neighbor C atoms were separated by 9.4 Å. The surface brillouin zone was sampled by a 6×6 equal-spaced set of k-vectors, including $\overline{\Gamma}$, incurring expected binding energy errors <0.01 eV, based on a spot check using a 9×9 sample. Electronic convergence was accelerated by means of Methfessel-Paxton smearing of the Fermi level (width = 0.2 eV) [40]. The Neugebauer-Scheffler method [41] was used to cancel unphysical fields resulting from the differences imposed on the upper and lower slab surfaces.

---

[1] While graphene nucleation occurs preferentially at Ru steps and step bunches, we have not found that Ru steps influence C segregation. That is, the same C adatom concentration is found for wide Ru terraces as narrow terraces.

## 3. Results and discussion

*3.1. Carbon adatom concentrations in equilibrium with bulk carbon*

Figure 1 illustrates how segregated C adatoms decrease the electron reflectivity of the Ru(0001) surface as a function of electron energy. At 1046°C, there are essentially no C adatoms on the surface (see discussion below and Fig. 3). At 846°C, however, C adatoms do exist and decrease the reflectivity, particularly of 3-6 eV-electrons. The reflectivity decrease caused by cooling is essentially identical to what we observe after depositing C on the surface using a C evaporator [28]. Thus, we are confident that the reflectivity change is caused by C segregating from the crystal. Indeed, Auger spectroscopy confirmed that C was abundant on surfaces for which graphene islands were observed by LEEM.

The C adatom concentration was determined from the reflectivity changes of 3.6-eV electrons, using the following procedure: first, the crystal was equilibrated at a temperature high enough that essentially no C was on the surface. Then the temperature was dropped and the reflectivity change monitored until the signal became constant, i.e., until equilibrium between the surface and bulk was achieved. Figure 2 shows how the adatom concentration increases with time after cooling. The equilibrium adatom concentration also increases with decreasing temperature. Below about 800°C, the adatom concentration eventually decreased when graphene nucleated on the surface, as discussed in section 3.2.

Figure 3a displays $c_s^{eq,b}$, the adatom concentration in equilibrium with C in Ru's bulk, as a function of temperature. The figure clearly shows that the crystal begins to segregate C below about 950°C, confirming our assertion that the surface is essentially C free at higher temperature. Figure 3a also shows a fit of the experimental data to the simplest treatment of surface segregation, the Langmuir-McLean model [11]:

$$\frac{c_s^{eq,b}}{1-c_s^{eq,b}} = \frac{c_b}{1-c_b} e^{-\Delta G_{seg}/kT} \qquad \text{eqn. 1,}$$

where $c_b$ is the C concentration in Ru, $\Delta G_{seg}$ is the free energy of segregation, $k$ is the Boltzmann constant, and $T$ is temperature. This model assumes that C forms an "ideal", non-interacting gas on a lattice of well-defined binding sites in both the bulk and on the surface. The slope of the Langmuir-McLean model plotted in an Arrhenius from (i.e., $\ln(c_s^{eq,b}/1-c_s^{eq,b})$ vs. $1/T$) gives the enthalpy of segregation: $\Delta H_{seg}$ = -2.4±0.2 eV. The dashed red line in Fig. 3c shows the reasonable fit of the data to the Langmuir-McLean model. The enthalpy of forming a C adatom from graphene on Ru(0001) is ~0.3 eV [28]. Using this value and the measured segregation enthalpy, we can estimate the formation enthalpy of dissolving a atom C from graphene on Ru into the bulk as 2.4 – 0.3 = 2.1 eV. (This value is small compared the cohesive energy of graphite, 7.4 eV [42].)

Table 1 shows the energy (enthalpy) cost to move a C adatom to various interstitial and substitutional sites inside the Ru lattice, as calculated (see section 2) from first-principles techniques. (The low-energy state for a C adatom is in an hcp hollow site [28].) The lowest energy sites for C interior to the surface are the octahedral interstitial sites. Moving a C adatom to these sites costs between 1.1 and 1.4 eV, with the higher energy corresponding to sites farther from the surface. Thus, the experimental segregation enthalpy is high compared to the value expected from the first-principles calculations.

The large disagreement suggests that the simple Langmuir-McLean model might be inappropriate. One should also be aware that the segregation energies obtained from the Langmuir-McLean model for certain other C/transition metal systems have been found to be greater than segregation energies obtained from the more accurate method of isosteres (i.e., the relationship between $c_b$ and $T$ at constant $c_s^{eq,b}$) [11] because of the effects of adatom-adatom interactions. (The isostere treatment contains fewer assumptions than the Langmuir-McLean model.) Another possibility for the discrepancy is that C atoms in Ru attract each other [43], resulting in carbon clusters in the bulk. Note that Potekhina et al. have measured C segregation energies as high as 2.3 eV on tungsten surfaces [44], a system that forms a bulk carbide.

Table 1. Calculated energies required to move a C atom from the hcp site on top of the slab to various interior sites.

| | |
|---|---|
| hcp C adatom → octahedral site between layers 1 and 2 | +1.1 eV |
| hcp C adatom → open-above tetrahedral site between layers 1 and 2 | < 0 (unstable, atom moves up) |
| hcp C adatom → under-Ru tetrahedral site between layers 1 and 2 | +1.6 eV |
| hcp C adatom → octahedral site between layers 3 and 4 | +1.4 eV |
| hcp C adatom → tetrahedral site between layers 3 and 4 | +2.7 eV |
| hcp C adatom → substitutional site in center of slab | +3.8 eV |

We emphasize that Fig. 3a shows the temperature dependence of the adatom concentration in equilibrium with a particular, *fixed* concentration of bulk C. That is, Fig. 3a is an "isoconcentrate," which shift with the bulk C concentration. For example, if $c_b$ is reduced, the isoconcentrate will shift to lower temperatures.

*3.2. Competition between two carbon reservoirs for equilibrating carbon adatoms*

At the lowest temperatures in Fig. 2 (801 and 733°C) the adatom concentration first increased upon cooling, but then decreased. We next show that graphene nucleation caused the decrease. No graphene islands nucleated within the LEEM field-of-view for the experiments in Fig. 2, but graphene was found in surrounding regions after completing the adatom concentration measurement. Figures 4 show an example where graphene nucleated within the field-of-view during cooling. The first island nucleated at 458 s. Figure 5a shows the concentration measured nearby, in a graphene-free region. (The box shown in Fig. 4.) The adatom concentration (see Fig. 5a) began to decrease at about 480 s, shortly after graphene nucleation. Clearly, the adatom concentration fell because graphene growth was consuming adatoms. Figure 3a also shows that the (metastable) maximum adatom concentrations observed before graphene nucleation (open squares) are smaller than the extrapolated equilibrium values.[2] That is, graphene nucleated before equilibrium between the C adatoms and the C in Ru was established.

---

[2] The large supersaturation needed to nucleate graphene [28] (see, for example, the 733°C dataset of Fig. 2) could potentially lead to hysteresis, particularly in the total amount of C on the surface. Our method of using imaging to know when graphene has nucleated and measuring the C adatom concentration independent of the graphene coverage avoids hysteretic effects.

Thus, C in the Ru/C system can exist in three states in our experiments: 1) dissolved in the bulk, 2) as adatoms, and 3) in a graphene sheet adjacent to the substrate.[3] Thus, the C adatoms can exchange with two reservoirs of condensed C (graphene and C in the Ru) that have different chemical potentials. The observation that the adatom concentration decreases when graphene nucleates clearly shows that there is a competition between the two reservoirs to equilibrate the adatoms.

We discuss this competition using the labeled schematic shown in Fig. 3b. Upon cooling from high temperature, the adatom concentration increases to maintain equilibrium with the bulk C (red curve in Fig. 3b). The adatom concentration decreases after graphene nucleates because the C in graphene has a lower chemical potential than the C in Ru. After graphene nucleates, the adatom concentration slowly decreases to the value in equilibrium with graphene. This approach to equilibrium is very slow, taking 100s of seconds in the 733 and 801°C datasets of Fig. 2. The adatom concentration does not easily come into equilibrium with the graphene because a large barrier impedes adatom attachment to the graphene [28]. The rate of equilibration of the adatoms is proportional to the graphene step length per unit surface area [28]. The graphene step length/area can vary greatly because the density of graphene islands that nucleate can vary greatly, depending mainly on the cooling rate and nucleation temperature. The kinetic aspect of the adatom concentration (also see section 3.3) is illustrated by comparing the 733 and 801°C datasets of Fig. 2. Compared to 801°C, the adatom concentration at 733°C quickly decreases from its maximum value, because a relatively high density of graphene islands nucleated. Graphene nucleation was very sparse at 801°C, giving little length of graphene step edge to capture the adatoms. Thus, the adatom concentration only slowly decreases toward the value for equilibrium with graphene.

---

[3] The fourth possible state is in graphene sheets not adjacent to the substrate. However, our Ru crystal contained insufficient C to precipitate multilayers.

For a given C concentration in Ru, there is only one temperature where the adatoms can be in equilibrium with both reservoirs of condensed C. This temperature is where the two equilibrium curves in Fig. 3 cross.[4] Above this temperature, there is no competition -- graphene is not stable and will dissolve. Below this temperature, however, achieving equilibrium is difficult. C will continue to segregate until the surface is covered with graphene. If the bulk C concentration is sufficiency high, additional graphene layers can grow and nucleate. But, given that the relative amount of C in the graphene is small compared to that in the bulk of the macroscopic sample, the segregation is not able to alter the bulk C concentration appreciably. Thus, the C in the Ru and the C in graphene cannot generally be in equilibrium.

*3.3. Kinetics of graphene growth from segregating carbon adatoms*

*3.3.1. Initial graphene growth rate from the supersaturated sea of C adatoms*

Figure 5 analyzes in detail the growth rate of the graphene island in Fig. 4. The essential measure of graphene growth rate is the velocity $v$ at which the edges of graphene sheets advance as a function of C adatom concentration [45]. For islands, the edge velocity is $v = (1/P)\,dA/dt$, where $P$ and $A$ are the island perimeter and area, respectively [28]. We measure both these quantities from LEEM images (see Fig. 5b) while simultaneously determining the C adatom concentration in a graphene-free region adjacent to the island. Fig. 5c shows that the graphene growth rate is a non-linear function of the adatom concentration.

---

[4] The two experimental equilibrium curves in Fig. 3a cross at a temperature slightly lower than where we observe graphene dissolution. This lack of accuracy likely results from the fact that the two curves were separately measured using different crystals and somewhat different apparatus and conditions (see section 2).

This result is entirely consistent with our previous study of graphene growing from vapor-deposited C [28]. There we concluded that growth involves the addition of rare clusters of about five atoms rather than of the much more abundant monomers (adatoms). These cluster-addition kinetics cause the non-linear dependence of growth rate on adatom concentration. The fact that we observe the same growth law for segregation as for vapor deposition strongly suggests that graphene growth during segregation also occurs primarily through the 2D gas phase. That is, the graphene growth mechanism is independent of the C source (from the bulk or vapor). However, graphene growth from segregating C is considerable less controllable than vapor-phase growth. Thus, we cannot study the segregation growth kinetics with the same level of detail as in vapor-phase deposition.

*3.3.2. Bulk diffusion controls long-term growth rate of graphene*

Clearly graphene grows as a non-linear function of the adatom concentration. But what controls the adatom concentration on the surface during segregation from the bulk? That is, what controls the overall graphene growth rate? Insight is gained from Fig. 5d, where we plot the square of island area vs. time. At times greater than ~ 650 s, the area squared grows linearly. We now show that this result proves that the long-term growth rate is controlled by the rate at which C diffuses through the Ru crystal [46].

To model the graphene growth rate, we make two assumptions. First, the adatom concentration quickly equilibrates with the bulk C concentration immediately below the surface. Second, the adatom concentration is constant with time. Support for the first assumption comes from oscillating the Ru temperature when no graphene is present and monitoring the adatom concentration. We find that temperature and adatom concentration always change in phase, even for the fastest oscillations we can impose.[5] From this result we conclude that the surface can quickly obtain the temperature-dependent adatom concentration needed for equilibrium with the bulk C concentration immediately below the surface.

Let the bulk concentration in the direction $z$, perpendicular to the surface, be $c_b(z,t)$, where $t$ is time. The time-dependent, one-dimensional diffusion equation is:

$$\frac{\partial c_b}{\partial t} = D_b \nabla^2 c_b \qquad \text{eqn. 2}$$

where $D_b$ is the bulk diffusion constant. For boundary conditions, our assumptions of a constant adatom concentration in equilibrium with the bulk C adjacent to the surface give $c_b(z=0,t) = c_0$. And, we take the initial bulk concentration to be uniform with depth: $c_b(z>0,t=0) = c_1$. Equation 1 then has solution

$$\frac{c_b(z,t) - c_0}{c_1 - c_0} = erf\left(\frac{z}{2\sqrt{D_b t}}\right) \qquad \text{eqn. 3}$$

where $erf$ is the error function. The overall graphene growth rate is proportional to the flux of C segregating to the surface, $D_b \nabla c_b|_{z=0}$. This flux is partitioned among the various graphene islands. Since our data come from well-separated graphene islands, we assume that the flux going to a given island is a constant described by the island's "capture area" $A_0$ [47]. Then the island area $A(t)$ is:

$$A(t) = A_0 \int_0^t D_b \nabla c_b|_{z=0} dt = \frac{A_0 D_b (c_1 - c_0)}{\sqrt{\pi D_b}} \int_0^t \frac{dt}{\sqrt{t}} = \frac{2 A_0 (c_1 - c_0)\sqrt{D_b t}}{\sqrt{\pi}} \qquad \text{eqn. 4.}$$

This model of bulk-diffusion-limited growth gives $A^2 \propto t$, as observed in Fig. 5d at later times.

Figure 5a shows that our second assumption is not strictly true; the adatom concentration was not constant during growth, but slowly decreased. Again, the slow response for the adatoms to come into equilibrium with the graphene results from the large barrier for adatoms to attach to graphene [28]. Thus, the bulk concentration adjacent to the surface is also not fixed at $c_0$, as assumed above. While not shown here, we have analyzed the effect of a time-dependent adatom concentration on the rate of surface segregation. Assuming linear growth kinetics, for simplicity, we find that the existence of an attachment barrier causes a small correction to the segregation flux. Still, the island area grows like eqn. 4, as $\sqrt{t}$. The physical picture is that the growth rate after the initial spurt is determined by diffusional gradients in the bulk. The adatom concentration adjusts to the value needed to overcome the graphene-edge barrier; so, the growth rate matches the segregation rate.

---

[5] The fastest temperature oscillations were about 40 s/period.

Additional insight into the interplay between growth rate and the driving forces for segregation and C incorporation into graphene comes from Fig. 6. This plot analyses the same island growing at two temperatures. At the nucleation temperature (818°C), the island area squared increases linearly in time. Increasing temperature causes the island to shrink slightly and then stop growing.[6] The island only begins to grow again after the adatom concentration increases. Eventually the area squared again increases linearly in time and the island grows at the segregation rate. But the growth rate is considerably slower than at the initial lower temperature (818°C). Even though bulk diffusion is faster at higher temperature (888°C), the driving force ($c_1 - c_0$ in eqn. 4) for segregation is smaller because the adatom concentration is closer to the value in equilibrium with the C in bulk Ru.

To summarize graphene growth from segregating C, if the adatom concentration becomes high enough during cooling, graphene islands nucleate and grow. Initial graphene growth is fed by depleting the adatom sea, which is highly supersaturated at nucleation. As this "store" of readily accessible surface C is depleted, growth slows. Then the graphene grows only at the rate at which bulk diffusion brings C to the surface.

---

[6] The shrinkage shows that the adatom concentration was temporarily below the value in equilibrium with graphene after the temperature increase. That is, the adatom concentration goes below the black curves in Fig. 3. The island shrinks until the adatom concentration comes into equilibrium with graphene.

**4. Summary and conclusions**

Using an imaging approach to measure carbon adatom concentrations, we have clarified qualitative and quantitative details about the classical surface-science phenomena of segregation. We measured the temperature-dependent adatom concentration in equilibrium with the C dissolved in bulk Ru (Figs. 2 and 3a). The data, combined with our previously determined measurements of the adatom concentrations in equilibrium with graphene (Fig. 3a), allow us to understand the competition that apportions C among three possible states, dissolved in Ru, as an adatom, or in graphene (Fig. 3b). We have found that graphene island growth always occurs from C atoms in the 2D surface gas surrounding the islands, rather than by direct attachment from the bulk. As for graphene growth from vapor-deposited C, the growth rate from segregating C is a non-linear function of the adatom supersaturation (Fig. 5c). Initially graphene grows rapidly by depleting the large supersaturation needed for nucleation (Fig. 5). However, over longer times, the graphene grows as fast as C segregates by bulk diffusion (Fig. 5d) and the coverage increases as $\sqrt{t}$. Our measured segregation enthalpy is higher than the value predicted by first-principles calculations. Some of the discrepancy may stem from not obtaining experimental numbers from isosteres. Some may be attributable to positional correlations/clustering of C atoms in bulk Ru.

**Acknowledgments**

The authors thank J. C. Hamilton for informative discussions. This work was supported by the Office of Basic Energy Sciences, Division of Materials Sciences and Engineering of the US DOE under Contract No. DE-AC04-94AL85000.

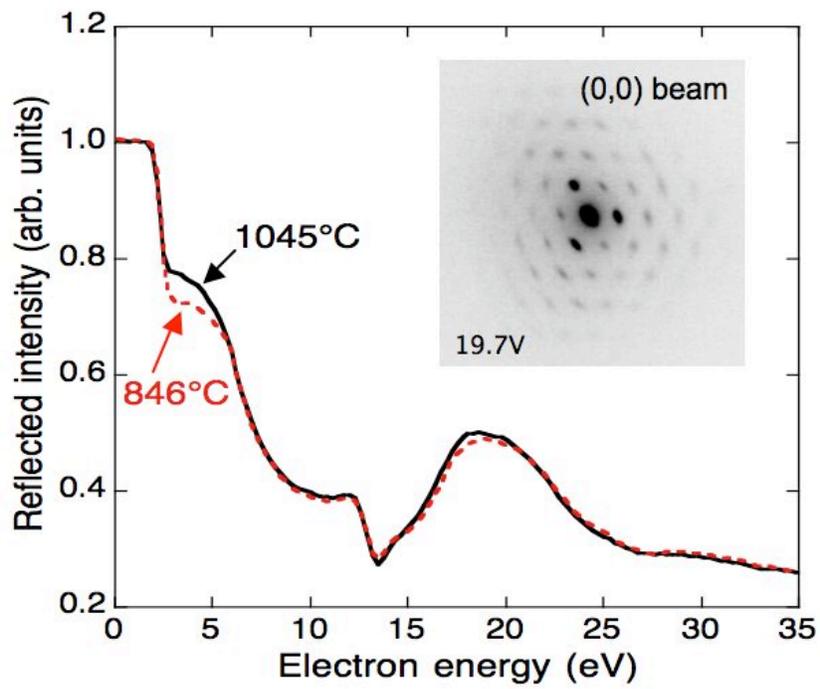

Fig. 1. Electron reflectivity versus electron energy from the Ru(0001) surface at two temperatures. At 1045°C, there are essentially no C adatoms. After cooling to 846°C, about 0.01 ML of mobile C adatoms have segregated from the Ru bulk to the surface. Insert: LEED pattern from epitaxial graphene on Ru(0001) showing the superstructure spots around the specular electron beam. The graphene was nucleated and grown from segregating C at 645 and 745°C, respectively.

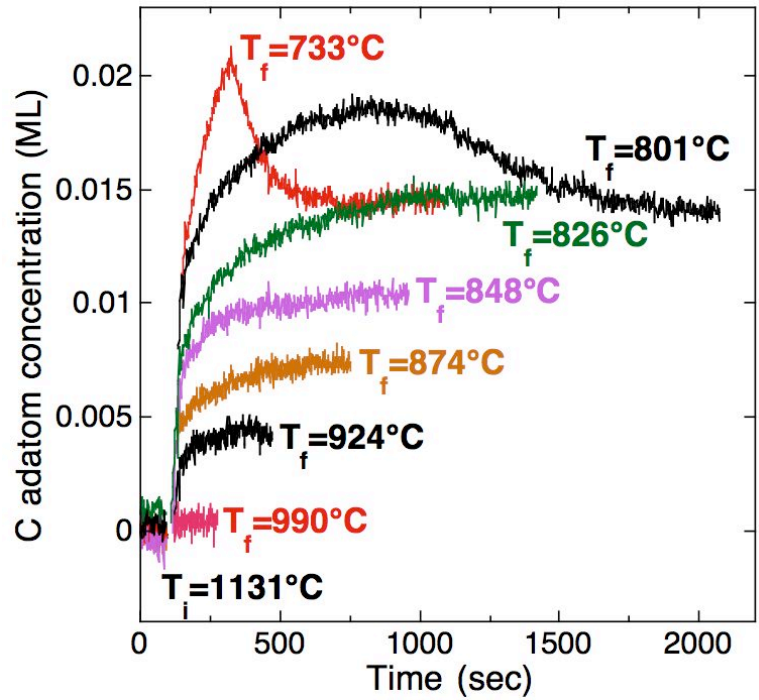

Fig. 2. a) Time evolution of the C adatom concentration during a temperature drop from $T_i$ = 1131°C (which is devoid of C adatoms) to the labeled temperatures, $T_f$. For $T_f$ < 826°C, the adatom concentration becomes sufficiently great for graphene islands to nucleate. Graphene growth reduces the adatom concentration.

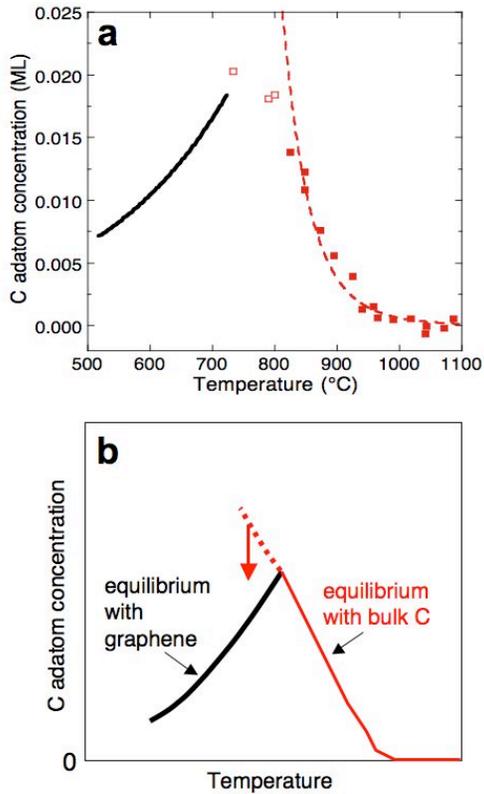

Fig. 3. a) Adsorption isoconcentrate showing C adatom concentrations (filled squares) in equilibrium with a fixed concentration of *bulk C in Ru*. Dashed red line shows fit to Langmuir-McLean segregation model (eqn. 1). Open squares show the maximum (metastable) concentrations observed before graphene nucleation (see Fig. 2). Black curve shows the adatom concentration in equilibrium with *graphene* [28]. b) Schematic illustration of the C adatom concentration in equilibrium with C in Ru's bulk (red curve) and in equilibrium with graphene (black curve). The dashed red curve shows the metastable adatom concentration in equilibrium with the bulk C but out-of-equilibrium with graphene.

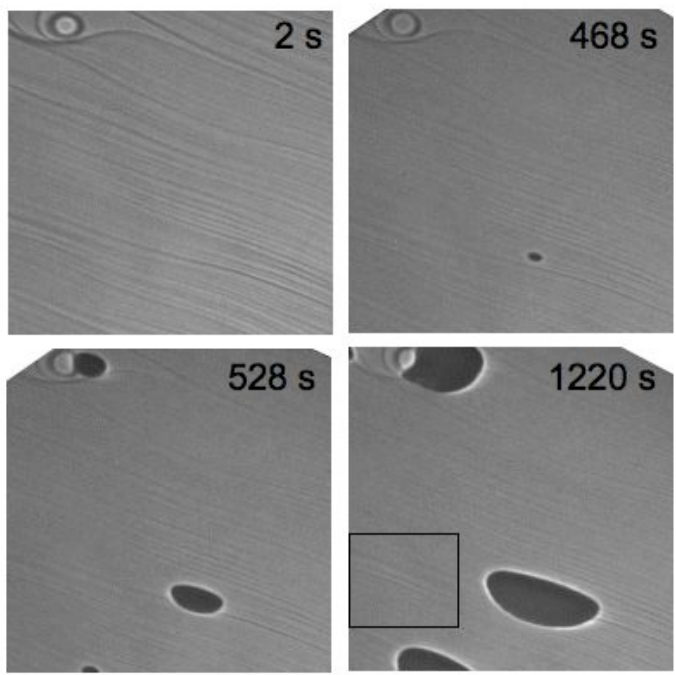

Fig. 4. A. Time sequence of LEEM images showing the nucleation and growth of graphene islands (black blobs) formed from C segregating from the Ru bulk. Field-of-view = 26 μm. Box in the last image shows the graphene-free region where the C adatom concentration was measured from the change in image intensity (see Fig. 5a).

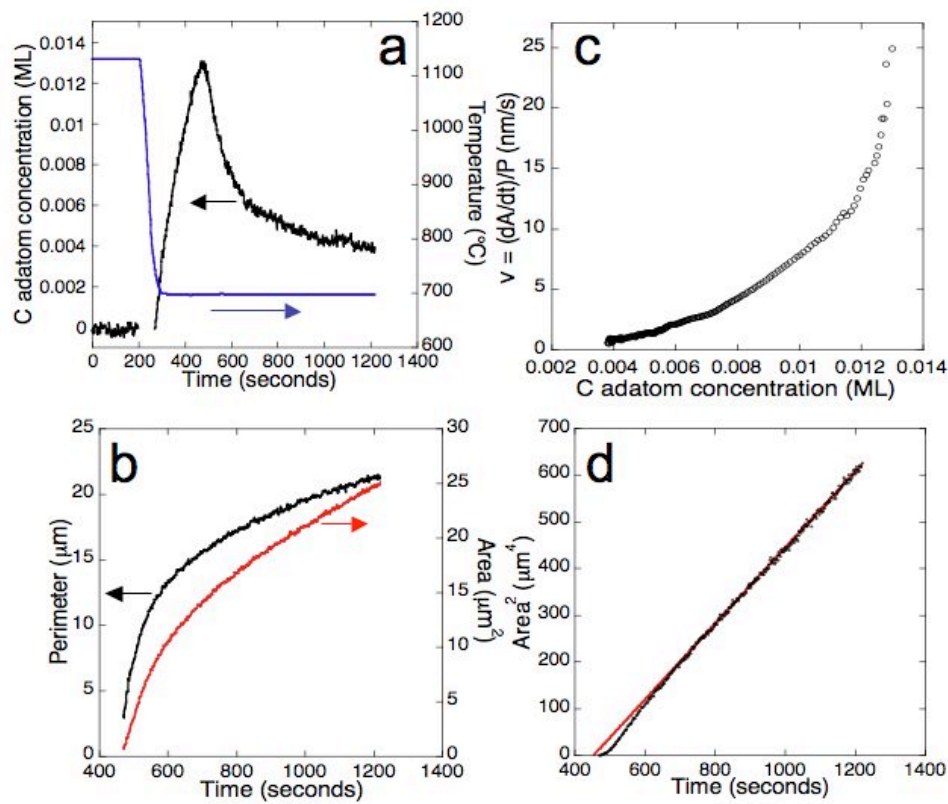

Fig. 5. Characterizing the growth rate of graphene island of Fig. 4. a) Ru temperature and C adatom concentration vs. time. b) Island area and perimeter. c) Velocity of graphene step edge as function of C adatom concentration. d) Square of island area vs. time. The red line is a linear fit to data for time >700 s.

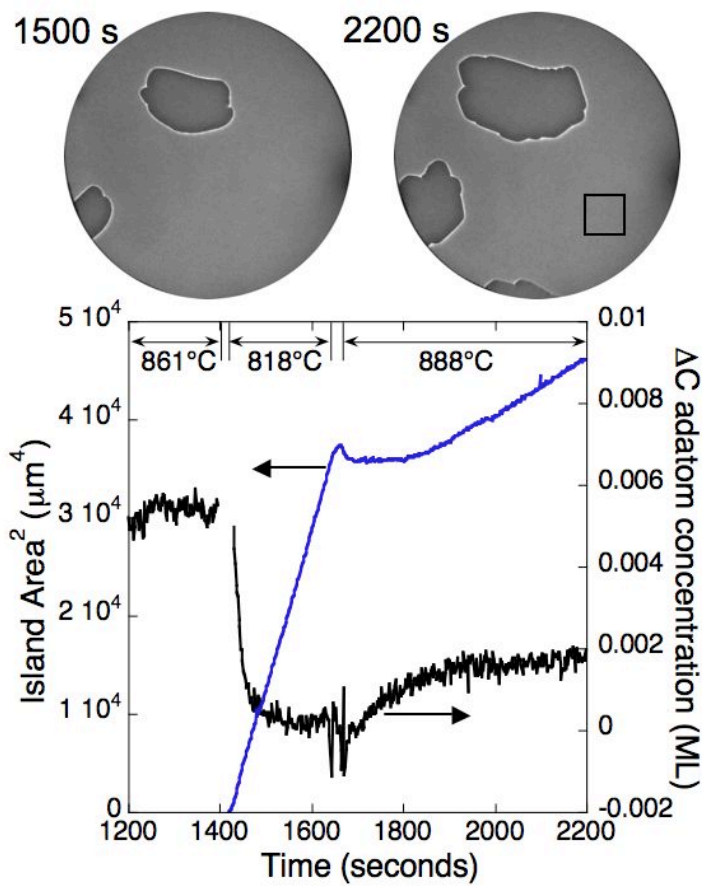

Fig. 6. Images: LEEM imaging of graphene island growing from segregating C. The island nucleated after cooling to 818°C. At about 1650 s, the temperature was increased to 888°C. Field-of-view is 45 μm. Box in the last image shows the graphene-free region where the C adatom concentration was measured from the change in image intensity. Plot: Square of island area and change in C adatom concentration during temperature excursions. The minimum adatom concentration in the plot is arbitrarily set to be 0 ML.